\documentclass[a4paper,11pt]{article}
\pdfoutput=1
\usepackage{jcappub}
\usepackage{amssymb,amsmath,mathrsfs,enumerate}
\usepackage{graphicx,rotate,multicol}
\usepackage{float}

\usepackage{subfig}

\usepackage{slashed}
\usepackage{mathtools}
\usepackage{multirow}

\allowdisplaybreaks

\title{\boldmath Spin of Primordial Black Holes from Broad Power Spectrum: Radiation Dominated Universe}

\author[a]{Indra Kumar Banerjee,}
\author[b]{Tomohiro Harada}
\affiliation[a]{Department of Physical Sciences, Indian Institute of Science Education and Research Berhampur,\\Transit Campus, Government ITI, Berhampur 760010, Odisha, India}
\affiliation[b]{Department of Physics, Rikkyo University, Toshima, Tokyo 171-8501, Japan}

\emailAdd{indrab@iiserbpr.ac.in}
\emailAdd{harada@rikkyo.ac.jp}

\abstract{We perform a study to obtain the initial spin or the nondimensional Kerr parameter $a_{*}$ of primordial black holes (PBHs) created during the radiation dominated phase of the universe from not only nearly monochromatic but also broad curvature power spectra. Motivated by inflation and first-order phase transitions, we consider a power law shape for the curvature perturbation. Although we can naturally neglect the contribution from the length scales smaller than the scale of interest, that from the larger scales may potentially be significant for a broad power spectrum, for which the spin is sensitive to the width of the power spectrum. So, we introduce a width parameter $r_{k}$, the ratio of the largest scale to the length of interest. We find that the root mean square of $a_{*}$ is largest for PBHs created from locally nearly scale invariant curvature power spectra with $r_{k} \sim 3.5$.
The upper limit is $\sim 1\times 10^{-4}$ for $M=10^{17}-10^{23}$ g, 
 $\sim 1.7\times 10^{-4}$ for $M=1-100 M_{\odot}$ and 
$\sim 2.5\times 10^{-3}$ even for an incredibly large mass of $M=10^{50}$ g.
}

\begin{document}
\maketitle
\flushbottom


\section{Introduction}
\label{sec:intro}
Primordial black hole (PBH) is a topic of great interest to the community as PBHs can partially or entirely play the role of dark matter~\cite{Carr:2009jm, Carr:2020gox, Carr:2016drx, Carr:2017jsz}. Furthermore, they can also be the source of some of the gravitational wave signals observed in LIGO~\cite{Nakamura:1997sm, Bird:2016dcv, Sasaki:2016jop, Clesse:2016vqa, Raidal:2017mfl} and they can be a pivotal part in the creation of supermassive black holes~\cite{Kohri:2014lza, Clesse:2015wea}. Furthermore, light PBHs might be one of the only ways to test new physics such as Hawking evaporation~\cite{Hawking:1975vcx}.

Many different mechanisms have been proposed for the creation of PBHs such as single and multi field inflation~\cite{Carr:1993aq,  Bullock:1996at, Saito:2008em, Kawai:2021edk, Kawai:2021bye, Choudhury:2023hvf, Randall:1995dj, Garcia-Bellido:2016dkw, Braglia:2020eai, Kawai:2022emp}, collapse of cosmic string loops~\cite{Hawking:1987bn, Borah:2023iqo}, first order phase transitions~\cite{Crawford:1982yz, Kawana:2021tde, Baker:2021nyl, Huang:2022him, Kawana:2022olo, Liu:2021svg, Gouttenoire:2023naa, Lewicki:2023ioy}, etc. Most, if not all, of these mechanisms revolve around the collapse of regions due to curvature perturbation of different shapes and magnitude. Generically, PBHs are hypothesized to have been created from collapse of horizon sized perturbations with mass close to that of the horizon at the time of creation. If, due to the curvature perturbation, there are overdense regions or density peaks, which are above a certain threshold value, then they can collapse to form PBHs~\cite{Carr:1975qj, Polnarev:2006aa, Musco:2008hv, Harada:2013epa, Harada:2015yda, Musco:2012au, Shibata:1999zs, Germani:2018jgr, Musco:2018rwt, Escriva:2019nsa, Escriva:2019phb, Yoo:2018kvb, Yoo:2020dkz}. However, in the near critical case where the density of peaks is very close to the threshold value, then PBHs may also form with mass much less than the horizon mass~\cite{Niemeyer:1999ak, Musco:2008hv, Musco:2012au}. 

Apart from the mass of the PBH, spin is also a very important property of PBHs as any uncharged black hole can be characterized by its mass and spin. The spin of PBHs have many important implications, as they can alter the Hawking evaporation spectra~\cite{Page:1976ki, Dasgupta:2019cae}, can leave their imprint in the gravitational waves created from their merger~\cite{Nakamura:1997sm}, can cause superradiant instability~\cite{Calza:2023rjt}, etc. Furthermore, PBH spin may also have changed throughout their lifetime due to Hawking evaporation~\cite{Arbey:2019jmj}, superradiance~\cite{Calza:2023rjt}, accretion~\cite{DeLuca:2020bjf} and other gravitational interactions such as merger~\cite{Barausse:2009uz} and close hyperbolic encounters~\cite{Jaraba:2021ces}. Therefore, in the last few years, many studies have been performed to obtain and understand the initial values of PBH spin~\cite{Chiba:2017rvs, Harada:2017fjm, DeLuca:2019buf, He:2019cdb, Mirbabayi:2019uph, Harada:2020pzb}. 

In this article, we follow the treatment of Ref.~\cite{Harada:2020pzb}, where spin has been characterized as a first-order effect of the perturbation.  This work focused on a nearly monochromatic power spectrum. However, in this study, we consider a broad power spectrum of the curvature perturbation with a power law dependence, where the `spread' of the curvature perturbation is characterized by the power law index and the width parameter. This is motivated from the fact that inflation and first order phase transitions give rise to nearly scale invariant and broad curvature perturbation, respectively. In our study, we remain in the domain of radiation dominated universe and obtain the dependence of the PBH spin on the power law index and the width parameter.

This article is organized as follows, in Sec.~\ref{sec:basdef}, we briefly review the mathematical foundation required for this study followed by Sec.~\ref{sec:bps}, where we describe the form of the power spectrum we consider and its consequences. In Sec.~\ref{sec:calcspin}, we obtain the spin and show the dependence of the spin on the power law index and width. We describe our results in Sec.~\ref{sec:res} and finally, in Sec.~\ref{sec:concl}, we summarize and conclude.
\section{Basic Definitions}
\label{sec:basdef}
In this section, we begin with some basic recap and definition of the formation of PBH and the angular momentum of the PBH. It is to be mentioned here that in this study we mainly follow the formalism of peak theory based on Ref.~\cite{1988heavens}, where the density and the curvature perturbation and the velocity of a region are considered as probability fields and we work with the correlations (of different orders) of these fields to fields to obtain the angular momentum and eventually the spin of the region which eventually collapses to form a PBH.

In this regard, we use the results from cosmological linear perturbation~\cite{Kodama:1984ziu}, where the solutions of the Einstein equation for gauge invariant perturbation quantities during radiation dominated universe can be expressed as~\cite{Harada:2020pzb}
\begin{align}
\Delta(x) &= D\sqrt{3}\left(\dfrac{\sin z}{z} - \cos z\right),\nonumber\\
V(x) &= D\left[\dfrac{3}{4}\left(\dfrac{2}{z^2}-1\right)\sin z-\dfrac{3}{2}\dfrac{\cos z}{z}\right],
\end{align}
where $D$ is an arbitrary constant, whose value depends on the shape of perturbation, $z=x/\sqrt{3}$, and $x=k\eta$.
For the CMC gauge, the density perturbation and the velocity of the region can be expressed as
\begin{align}
\delta_{\mathrm{CMC}} &= D\dfrac{\sqrt{3}z^2}{z^2+2}\left(2\dfrac{\sin z}{z}-\cos z\right), \nonumber\\
v_{\mathrm{CMC}} &= -\dfrac{3}{4}D\dfrac{(z^2-2)\sin z+2z\cos z}{z^2+2}.
\end{align}
For the conformal Newtonian gauge, the quantities take the form
\begin{align}
\delta_{\mathrm{CN}} &= \sqrt{3}D\dfrac{2(z^2-1)\sin z+(2-z^2)z\cos z}{z^4},\nonumber\\
v_{\mathrm{CN}} &= \dfrac{3}{4}D\dfrac{(2-z^2)\sin z - 2z\cos z}{z^2}.
\end{align}
It is to be noted here that we have only written the expressions of the necessary quantities here. For further insight on this matter, we recommend the reader to refer to Ref.~\cite{Harada:2020pzb} (and references therein).

Next, for the region which will eventually collapse to form a PBH, we denote the angular momentum in the region as $S_i$. Furthermore, the root mean square (RMS) value of the angular momentum can be expressed as
\begin{align}
\sqrt{\langle S_iS^i \rangle} = S_{\mathrm{ref}}(\eta)\sqrt{\langle s_{ei}s^i_e \rangle},
\end{align}
where $S_{\mathrm{ref}}$ is the reference angular momentum, which is common feature of all the regions with peak, and $s_e$ depends on the shape and the height of the peak. The reference angular momentum can be further expressed as~\cite{DeLuca:2019buf, Harada:2020pzb}
\begin{align}
    S_{\mathrm{ref}}(\eta)=\frac{4}{3}\left[a^4\rho_bg_{\mathrm{CN}}\right]_{\eta}(1-f)^{5/2}R_{*}^5,
\end{align}
where the quantities in the above expressions are defined as
\begin{align}
    R_{*}&=\sqrt{3}\dfrac{\sigma_1}{\sigma_2}\nonumber\\
    g_{\mathrm{CN}}^2(\eta_{\mathrm{ta}})&=\dfrac{4}{9}\int^{k_{\mathrm{max}}}_{k_{\mathrm{min}}}dk k^2 T_{v_{\mathrm{CN}}}(k,\eta_{\mathrm{ta}})\mathcal{P}(k)\nonumber\\
    T_{v_{\mathrm{CN}}}(k,\eta)&=\dfrac{\sqrt{3}}{8}\dfrac{(z^2-2)\sin z+2z\cos z}{z^2}.
\end{align}
Here, $T_{v_{\mathrm{CN}}}$ is the velocity transfer function in the comoving Newtonian gauge and $\sigma_i$ is the $i$th spectral moment of the density perturbation, which is explained in detail in the next section.
Furthermore, the RMS of $s_e$ can be expressed as~\cite{DeLuca:2019buf, Harada:2020pzb}
\begin{align}
\sqrt{\langle s_e^2\rangle} = 5.96 \times \dfrac{\sqrt{1-\gamma^2}}{\gamma^6\nu},
\end{align}
where $\nu$ signifies the height of the peak and $\gamma$ is an indicator of the width of the density perturbation. It is to be noted that both these quantities are explained in detail in the later part of this article. 

Finally, we can express the RMS value of the dimensionless Kerr parameter $a_*$ as~\cite{DeLuca:2019buf, Harada:2020pzb},
\begin{align}
\sqrt{\langle a_*^2\rangle} = A_{\mathrm{ref}}(\eta_\mathrm{ta})\sqrt{\langle s_e^2\rangle},
\end{align}
where $\eta_{\mathrm{ta}}$ is the point where the region decouples from the background (turn around point) and it is explained in detail in the later part of the article, and $A_{\mathrm{ref}}=S_{\mathrm{ref}}/(GM_{\mathrm{ta}}^2)$, where $M_{\mathrm{ta}}$ is the mass of the region at turn around point. Now with the foundation for the calculations laid, we proceed towards the treatment of a broad curvature power spectrum.
\section{Broad Power Spectrum}
\label{sec:bps}
The curvature perturbation we use in this work is as follows
\begin{align}
\mathcal{P}(k)=A^2(kR_{\mathcal{H}})^j,
\end{align}
where $j\in(0,\infty)$ and $R_{\mathcal{H}}$ is the comoving Hubble radius. It is to be noted here that $j\rightarrow\infty$ implies a monochromatic power spectrum, whereas $j\rightarrow 0$ implies a scale invariant power spectrum. It is also worth mentioning here that $A$ signifies the magnitude of the curvature perturbation.
\subsection{Spectral Moments}
\label{subsec:specmom}
A key ingredient in calculation of the PBH properties is the spectral moments of the density power spectrum at different order which is usually expressed as
\begin{align}
\sigma_n^2 = \dfrac{4}{9}\eta_{\mathrm{init}}^4\int^{\infty}_{0}\dfrac{dk}{k} k^{2n+4}\mathcal{P}(k).
\end{align}
However since in our case we work with broad spectrum, we do not consider contributions from length scales smaller than the horizon length, which makes the upper limit of the above integration $k_{\mathrm{max}}=(r_{\mathrm{max}})^{-1}=\mathcal{H}$. Furthermore, we find that as $j\rightarrow 0$, if the lower limit of the integration is 0, that lead to unphysical singularities. Therefore, we fix the lower limit of the integration as $k_{\mathrm{min}}$. Therefore, we can express the spectral moments as
\begin{align}
\sigma_n^2 &= \dfrac{4}{9}\eta_{\mathrm{init}}^4\int^{k_{\mathrm{max}}}_{k_{\mathrm{min}}}\dfrac{dk}{k} k^{2n+4}\mathcal{P}(k)\nonumber \\
&=\dfrac{4}{9}\eta_{\mathrm{init}}^4\int^{k_{\mathrm{max}}}_{\frac{k_{\mathrm{max}}}{r_k}}\dfrac{dk}{k} k^{2n+4}\mathcal{P}(k),
\end{align}
where $r_k=k_{\mathrm{max}}/k_{\mathrm{min}}$ is the ratio of the lowest and the highest length scales we consider and it signifies the width of the power spectrum. It is to be noted here that the value of $r_k$ depends on the specific mechanisms which create the perturbation. However, in this work, we keep this as a free parameter. It is to be noted here that the limits we impose on the integral essentially work as a `window' and hence we do not consider a window function in this work.
Finally, after performing the integration, we can express the spectral moments in a simplified form as
\begin{align}
\sigma_{n}^2 = \dfrac{4}{9}\eta_{\mathrm{init}}^4 A^2 k_{\mathrm{max}}^{4+2n}\dfrac{1-r_k^{-4-2n-j}}{4+2n+j}.
\end{align}
We have shown in the later part of the article that the initial spin of PBHs does not depend on $r_k$ for $j\gtrsim 0.1$, but it depends on $j$. However, for $j\lesssim 0.1$, it is opposite, i.e., the initial spin depends on $r_{k}$ but not on $j$. Therefore, in this article, we divide our parameter space in two different parts, i.e., (i) $j>0.1$, (ii) $j\leq 0.1$. In the first case, we take $r_k\rightarrow\infty$, (which signifies maximum width, i.e., $k$ goes from 0 to $k_{\mathrm{max}}$), which leads to an even more simplified form of spectral moments, i.e.,
\begin{align}
\sigma_{n}^2 = \dfrac{4}{9}\eta_{\mathrm{init}}^4 A^2 k_{\mathrm{max}}^{4+2n}\dfrac{1}{4+2n+j}.
\end{align}
For the case where $j\leq 0.1$, we consider $r_k\in(1,\infty)$. The value of $r_k$ has very important implications, which we will see in the subsequent parts of the article.
\subsection{Density and Curvature Profile}
\label{subsec:denprof}
The next task is to determine the shape of the density profile, which allows us to gain insight into the spatial behavior of the density perturbation. 

According to Ref.~\cite{Yoo:2018kvb}, the curvature profile can be expressed as\footnote{It should be noted here in Ref.~\cite{Yoo:2018kvb}, the authors only used symbols such as $g,~g_0,~g_1$ as they were only focusing on curvature perturbation.}
\begin{align}
g_{\zeta}(r;k_*) = g_{0, \zeta}(r) + k_*^2 g_{1, \zeta}(r),
\end{align}
where
\begin{align}
g_{0, \zeta} &= -\dfrac{1}{1-\gamma^{\prime~2}}\left(\psi_{\zeta}^2 + \frac{1}{3}R^{\prime}_*\Delta \psi_{\zeta}\right)\nonumber\\
g_{1, \zeta} &= -\dfrac{1}{\gamma^{\prime}(1-\gamma^{\prime~2})}\dfrac{\sigma^{\prime}_0}{\sigma^{\prime}_2}\left(\gamma^{\prime~2}\psi_{\zeta}^2 + \frac{1}{3}R^{\prime}_*\Delta \psi_{\zeta}\right),\nonumber\\
\psi_{\zeta}(r) &= \dfrac{1}{\sigma^{\prime~2}_0}\int\dfrac{dk}{k}\dfrac{\sin(kr)}{kr}\mathcal{P}(k),\nonumber\\
\gamma^{\prime} &= \dfrac{\sigma^{\prime~2}_1}{\sigma^{\prime}_0\sigma^{\prime}_2},\nonumber\\
R^{\prime}_* &= \sqrt{3}\dfrac{\sigma^{\prime}_1}{\sigma^{\prime}_0},\nonumber\\
\label{prof1}
\end{align}
where 
\begin{align}
\sigma^{\prime~2}_{n}&=\int_{k_{\mathrm{min}}}^{k_{\mathrm{max}}}\dfrac{dk}{k}k^{2n}\mathcal{P}(k).
\label{sigmaprime}
\end{align}
We get the profile for curvature perturbation as
\begin{align}
g_{\zeta}(r;k_*) = a_{\zeta}^* - b_{\zeta}^*(k_{\mathrm{max}}r)^2,
\end{align}
where
\begin{footnotesize}
\begin{align}
a_{\zeta}^* &= \frac{1}{1-\frac{j(j+4)(1-r_k^{-j-2})^2}{(j+2)^2(1-r_k^{-j})(1-r_k^{-j-4})}}\left(1-\frac{j(j+4)(1-r_k^{-j-2})^2}{3(j+2)^2(1-r_k^{-j})(1-r_k^{-j-4})}-\alpha_{*}\frac{2(j+4)(1-r_k^{-j-2})}{3(j+2)(1-r_k^{-j-4})}\right),\\
b_{\zeta}^* &= \frac{1}{1-\frac{j(j+4)(1-r_k^{-j-2})^2}{(j+2)^2(1-r_k^{-j})(1-r_k^{-j-4})}}\left(\alpha_*\frac{j(j+4)(1-r_k^{-j-2})^2}{6(j+2)^2(1-r_k^{-j})(1-r_k^{-j-4})}-\frac{j(1-r_k^{-j-2})}{6(j+2)(1-r_k^{-j})}\right),
\end{align}
\end{footnotesize}
and $\alpha_{*}=k_*^2/k_{\mathrm{max}}^2$, where $1/k_*$ is the length scale of the curvature peak.
Similarly, we can obtain the density profile as well using the above formalism by using the density spectrum, which can be expressed as
\begin{align}
\mathcal{P}_{\delta}(k) = \dfrac{4}{9}(kR_{\mathcal{H}})^4\mathcal{P}(k) = \dfrac{4}{9}A^2(kR_{\mathcal{H}})^{j+4}.
\end{align}
We obtain the profile for the density perturbation as
\begin{align}
g_{\delta}(r;k_*) = a_{\delta}^* - b_{\delta}^*(k_{\mathrm{max}}r)^2,
\end{align}
where
\begin{footnotesize}
\begin{align}
a_{\delta}^* &= \frac{1}{1-\frac{(j+4)(j+8)(1-r_k^{-j-6})^2}{(j+6)^2(1-r_k^{-j-4})(1-r_k^{-j-8})}}\left(1-\frac{(j+4)(j+8)(1-r_k^{-j-6})^2}{3(j+6)^2(1-r_k^{-j-4})(1-r_k^{-j-8})}-\alpha_{\delta}\frac{2(j+8)(1-r_k^{-j-6})}{3(j+6)(1-r_k^{-j-8})}\right),\\
b_{\delta}^* &= \frac{1}{1-\frac{(j+4)(j+8)(1-r_k^{-j-6})^2}{(j+6)^2(1-r_k^{-j-4})(1-r_k^{-j-8})}}\left(\alpha_{\delta}\frac{(j+4)(j+8)(1-r_k^{-j-6})^2}{6(j+6)^2(1-r_k^{-j-4})(1-r_k^{-j-8})}-\frac{(j+4)(1-r_k^{-j-6})}{6(j+6)(1-r_k^{-j-4})}\right),
\end{align}
\end{footnotesize}
and $\alpha_{\delta}=k_{\delta}^2/k_{\mathrm{max}}^2$, where $1/k_{\delta}$ is the length scale of the density peak. It is to be noted here that for both curvature and density profile, we only keep the first two terms of the Taylor expansion.
As mentioned in Ref.~\cite{Yoo:2018kvb}, we can express the curvature profile at the average value of $k_{*}=k_c=\sigma^{\prime}_1/\sigma^{\prime}_0$ as
\begin{align}
g_{\zeta}(r;k_c) = \psi_{\zeta}(r) = 1 - \dfrac{j(1-r_k^{-j-2})}{6(j+2)(1-r_k^{-j})}(k_{\mathrm{max}}r)^2.
\end{align}
Similarly, for the density perturbation at the average value of $k_{\delta} = k_{c\delta} = \sigma_1/\sigma_0$, the profile can be expressed as
\begin{align}
g_{\delta}(r;k_{c\delta}) = \psi_{\delta}(r) = 1 - \dfrac{(j+4)(1-r_k^{-j-6})}{6(j+6)(1-r_k^{-j-4})}(k_{\mathrm{max}}r)^2.
\end{align}
As mentioned before for the case where $j>0.1$, we use $r_k\rightarrow\infty$, which results in simplified forms of these profiles, i.e.,
\begin{align}
\psi_{\zeta}(r) &= 1-\left(\dfrac{j}{6(j+2)}\right)(k_{\mathrm{max}}r)^2,\\
\psi_{\delta}(r) &= 1-\left(\dfrac{(j+4)}{6(j+6)}\right)(k_{\mathrm{max}}r)^2.
\end{align}
It is worth mentioning here that if we consider $j\rightarrow\infty$, then the profile takes the form
\begin{align}
\psi_{\delta}(r) = \psi_{\zeta}(r) = 1-\dfrac{(k_{\mathrm{max}}r)^2}{6},
\end{align}
which is consistent with the shape of the profile for a monochromatic power spectrum. On the other hand, in the case where $j\rightarrow 0$ (keeping $r_k$ finite), the profiles take the form
\begin{align}
\psi_{\zeta}(r) &= 1-\dfrac{1}{12}\dfrac{1-r_k^{-2}}{\ln(r_k)}(k_{\mathrm{max}}r)^2,
\label{psizeta}
\end{align}
\begin{align}
\psi_{\delta}(r) &= 1-\dfrac{1}{9}\dfrac{(1-r_k^{-6})}{(1-r_k^{-4})}(k_{\mathrm{max}}r)^2.
\label{psidelta}
\end{align}
This is due to the fact that in the limit $j\rightarrow 0$, $(1-r_k^{-j})/j$ takes the form $\ln(r_k)$. Therefore, as mentioned before, to take the limit of scale-invariant perturbation, finite values of $r_k$ is crucial as otherwise the curvature profile loses its dependence on $r$ and $k_{\mathrm{max}}$, which is unphysical. As mentioned before, $r_k$ is a very important quantity, which signifies the window of contribution for nearly scale-invariant power spectra. To illustrate this further, if we consider $r_k\rightarrow 1$ in the Eqs.~\eqref{psizeta} and \eqref{psidelta}, the profile takes the form
\begin{align}
\psi_{\zeta}(r)= \psi_{\delta}(r) = 1-\dfrac{(k_{\mathrm{max}}r)^2}{6},
\end{align}
which is equivalent to the monochromatic power spectrum. This is expected as $r_k\rightarrow 1$ essentially mandates that we consider only the contributions from horizon length scale.
\subsection{Curvature and Compaction}
\label{subsec:curvcomp}
The profiles allow us to express the density and the curvature perturbations as~\cite{Harada:2020pzb}
\begin{align}
\delta_{\mathrm{CMC}}(\eta, \mathbf{r})&=\delta_{pk}(\eta)g_{\zeta}(r;k_*),\nonumber \\
\zeta(\eta,\mathbf{r})&=\zeta_{pk}(\eta)g_{\delta}(r;k_{\delta}),
\end{align}
where `$pk$' corresponds to the maximum value at a given $\eta$. It can be seen from the previous subsection that the density profile in case of $k_{\delta} = k_{c\delta}$ can be expressed as
\begin{align}
\psi_{\delta}(r) = 1 - \dfrac{(j+4)(1-r_k^{-j-6})}{6(j+6)(1-r_k^{-j-4})}(k_{\mathrm{max}}r)^2 = 1- \dfrac{(rk_{c\delta})^2}{6}.
\end{align}
This is similar to the truncated sinc-type profile considered in Ref.~\cite{Harada:2020pzb}. Therefore, following their footsteps, we identify $\delta_{pk}$ as $\delta_{\mathrm{CMC},k_{c\delta}}$ and we express it as
\begin{align}
\delta_{\mathrm{CMC},k_{c\delta}}(\eta)=\dfrac{2}{3}x^2(-\zeta_{k_{c\delta}}(0)).
\end{align}
Furthermore, the arbitrary constant in the expressions of the gauge-invariant perturbations mentioned in Sec.~\ref{sec:basdef} can be expressed as
\begin{align}
D=\dfrac{4\sqrt{3}}{3}(-\zeta_{k_{c\delta}}(0)).
\end{align}
Now we express the initial density perturbation in the CMC gauge to be
\begin{align}
\delta_{\mathrm{CMC},k_{c\delta}}(\eta_{\mathrm{init}},r) = \delta_{\mathrm{CMC},k_{c\delta}}(\eta_{\mathrm{init}})\psi_{\delta}(r).
\end{align}
Now our goal is to ensure that our method is consistent with the existing literature. In order to achieve that we obtain the expression for the compaction function. The compaction function is a measure of the amplitude of the curvature or density perturbation, i.e., it allows us to determine whether the perturbation is large enough for a collapse. We obtain the compaction function to be
\begin{align}
C(r) &=\left(\dfrac{\delta M}{ar}\right)(\eta_{\mathrm{init}},r)\nonumber\\ 
     &=\dfrac{1}{3}(k_{c\delta}r)^2\left(1-\dfrac{1}{10}(k_{c\delta}r)^2\right)(-\zeta_{k_{c\delta}}(0)).
\end{align}
We maximize the compaction function with respect to $r$ to find
\begin{align}
C_{\mathrm{max}} = C(r_{m}) = \dfrac{5}{6}(-\zeta_{k_{c\delta}}(0)),
\label{eq:compfunc}
\end{align}
where
\begin{align}
r_m = \sqrt{5}k_{c\delta}^{-1}.
\end{align}
The maximum value of the compaction function in radiation dominated universe is $\sim 2/5$~\cite{Shibata:1999zs, Harada:2015yda}. Comparing this value with Eq.~\eqref{eq:compfunc} we find
\begin{align}
\zeta_{k_{c\delta}}(0) &= -\dfrac{12}{25},\\
D &= \dfrac{16\sqrt{3}}{25}.
\end{align}
Now we use this value of $\zeta_{k_{c\delta}}(0)$ to obtain the density perturbation averaged over the overdense region. In order to obtain the `boundary' of the overdense region $r_0$, we equate the profile to zero, i.e., $\psi_{\delta}(r) = 0$, which gives us, $r_0 = \sqrt{6}k_{c\delta}^{-1}$. Now in the long wavelength limit, at horizon entry ($\eta_H = \eta_{\mathrm{init}}$), we consider $(aH)(\eta_H)r_0 = 1$ and obtain the average density perturbation as
\begin{align}
\overline{\delta}_H = \dfrac{2}{5} \dfrac{2}{3} (k_{c\delta}r_0)^2(-\zeta_{k_{c\delta}}(0)) = \dfrac{96}{125} = 0.768.
\end{align}
It is to be noted here that this value is within the range $(0.63,0.84)$, which were obtained in various studies~\cite{Polnarev:2006aa, Harada:2015yda}. Furthermore, it can be seen that the average value of the perturbation at $k_{\delta} = k_{c\delta}$ is independent from $j$ and $r_k$, which is also consistent with the fact that the perturbation threshold is property of the constituent of the universe at the time of collapse, i.e., it is governed by the equation of state parameter.
\section{Calculation of Spin}
\label{sec:calcspin}
In the previous section, we have calculated the necessary tools to calculate the spin of the PBH, which is the focus of this section.
\subsection{Turn Around Point}
\label{subsec:tap}
\begin{figure}[H]
\centering
\includegraphics[scale=0.35]{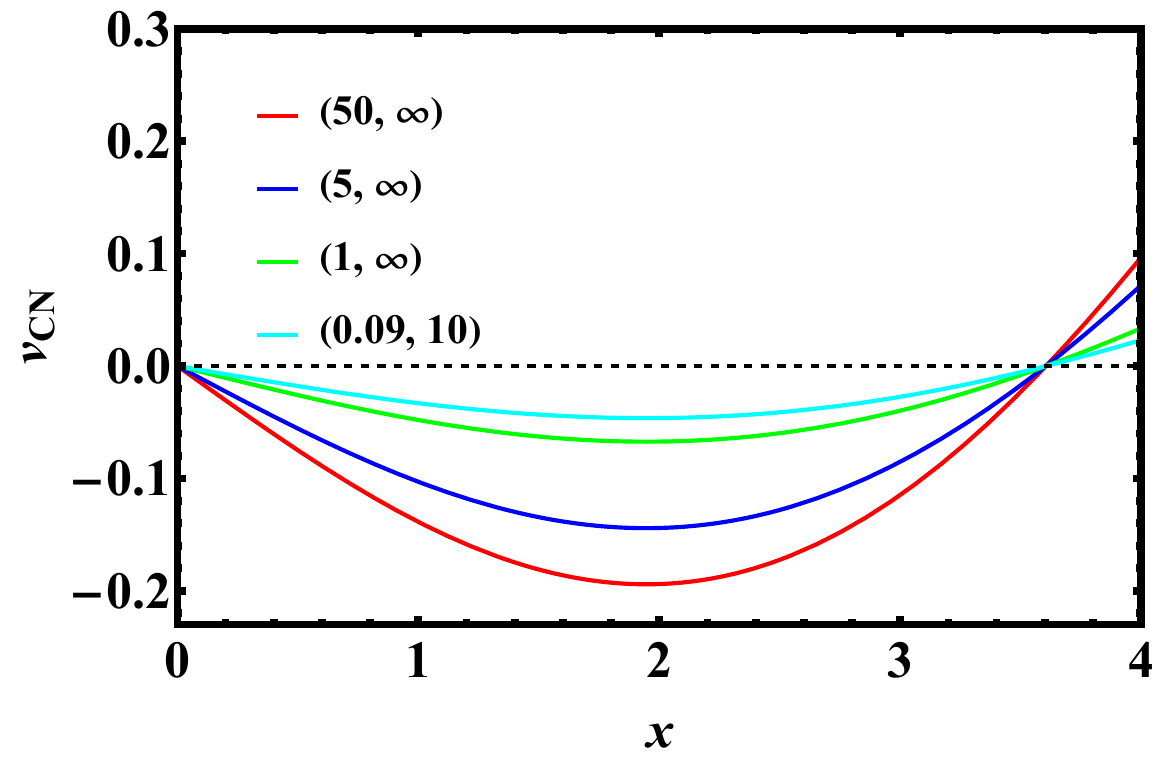}
\includegraphics[scale=0.35]{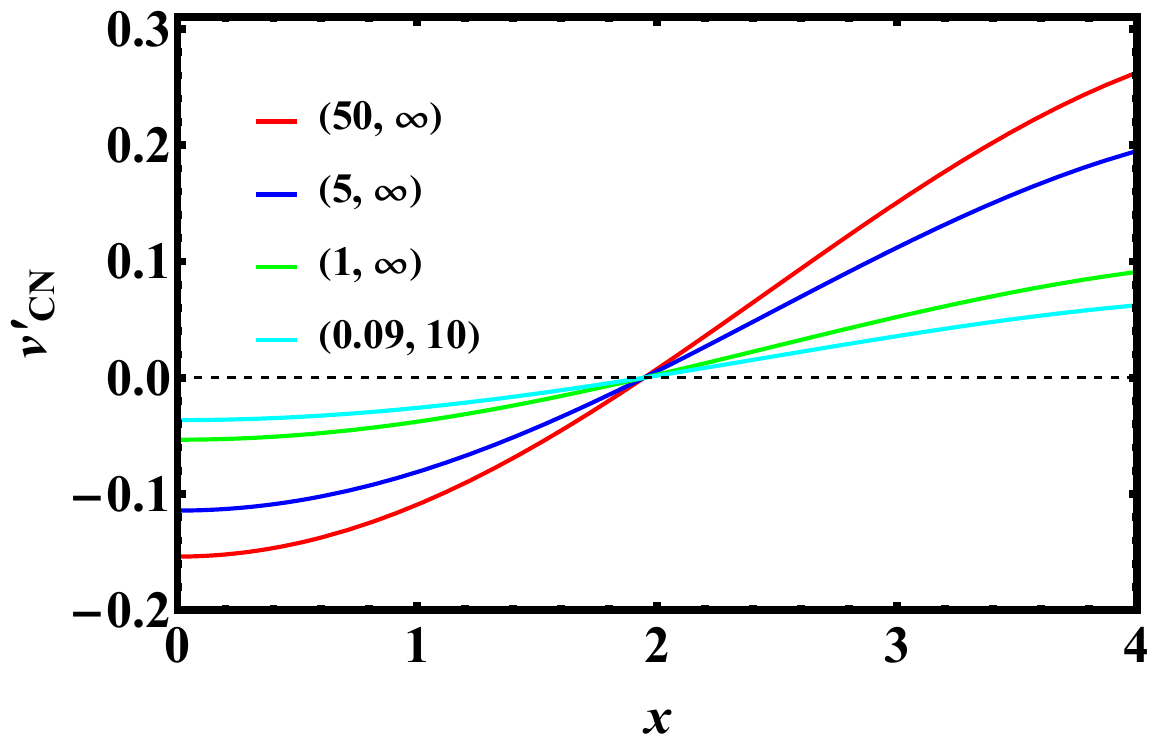}
\caption{We have shown the functions (Left) $v_{\mathrm{CN}}$ and (Right) $v^{\prime}_{\mathrm{CN}}$ as functions of $x=\eta k$. The different colors signify different values of $(j,~r_k)$, which are given in the plot legends.}
\label{velocity}
\end{figure}

One of the most important aspect of the calculation of the spin lies in the calculation of the `point' where the perturbation goes through a turn around, i.e., where a region's perturbation is so large that it decouples from the background to head towards the eventual collapse. However, determining this point is not an easy task. There are several criterion mentioned in literature where the turn around occurs. Following the footstep of Ref.~\cite{Saito:2023fpt}, we consider the criterion where the curvature perturbation is so large that it gets gravitationally attracted and the acceleration becomes inward. In a mathematical aspect, this can be expressed as the time when the velocity of the region reaches its minima. It can be considered that after this point in time, the perturbation is so large that it enters the regime of non-linearity. Therefore, $v^{\prime}_{\mathrm{CN}}(\eta_{\mathrm{ta}})=0$. In Fig.~\ref{velocity}, we show the the velocity of the region in the conformal Newtonian gauge and its derivative with respect to $x=\eta k$ at different values of $j$ and $r_k$.
It can bee seen from the figure that for all these different cases, the minima of the velocity occur at $x_{\mathrm{ta}}=1.95$, i.e., the value of the turn-around point does not depend on $j$ and $r_k$. This is due to the fact that although the magnitude of $v_{\mathrm{CN}}$ depends on $j$ and $r_k$ through the presence of $\Theta_{jk}$, the behavior depends only on $x$. Furthermore, for all these different cases, i.e., $(j,~r_k)~=~(50,\infty),~(5,\infty),~(1,\infty)$, and $(0.09,10)$, the values of the velocity at $x_{\mathrm{ta}}$ are $-0.193$, $-0.145$, $-0.069$, and $-0.048$.  
\subsection{Reference and RMS Spin}
\label{subsec:refrmsspin}
\begin{figure}[H]
\centering
\includegraphics[scale=0.355]{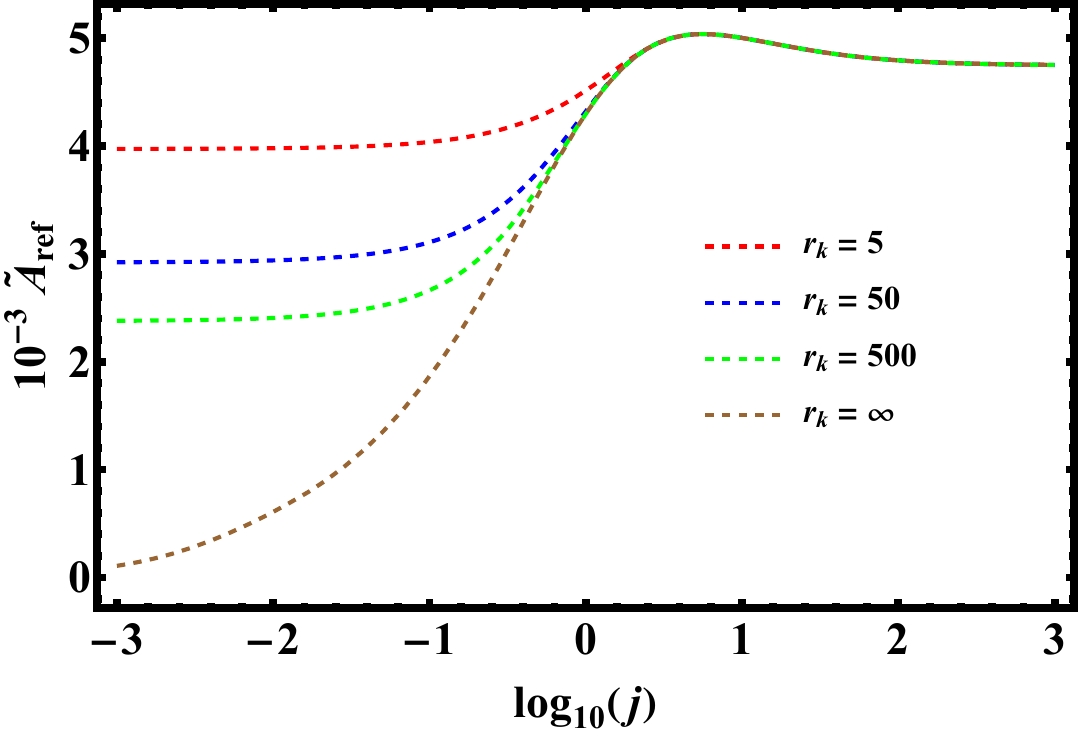}
\includegraphics[scale=0.36]{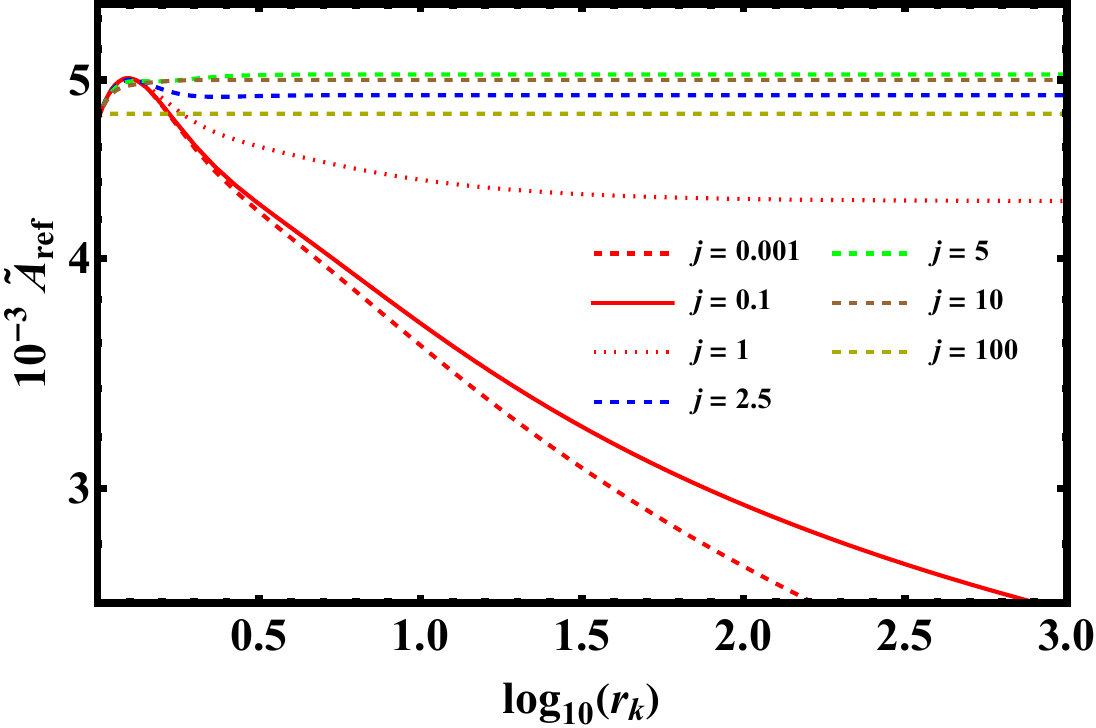}
\caption{(Left) The dependence of $\tilde{A}_{\mathrm{ref}}$ on $j$ for a few values of $r_{k}$. (Right) The dependence of $\tilde{A}_{\mathrm{ref}}$ on $r_k$ for a few values of $j$.}
\label{Areftilde}
\end{figure}

Now finally, we turn our focus to the spin of the region in question. The reference spin of the region at the turn around point can be expressed as
\begin{align}
    A_{\mathrm{ref}}(\eta_{\mathrm{ta}})=\dfrac{\frac{4}{3}\left[a^4\rho_bg_{\mathrm{CN}}\right]_{\eta=\eta_{\mathrm{ta}}}(1-f)^{5/2}R_{*}^5}{GM_{ta}^2}.
\end{align}
Furthermore, we can express $g_{\mathrm{CN}}(\eta)$ in the form
\begin{align}
    g_{\mathrm{CN}}(\eta)=\dfrac{2}{3}A k_{\mathrm{max}}G(\eta),
\end{align}
where
\begin{align}
    G^2(\eta)=\int^1_{1/r_k}dxx T_{v_{\mathrm{CN}}}^2(k,\eta)x^j.
\end{align}
Now, we can express $A_{\mathrm{ref}}$ as a function of both $j$ and $r_k$ in the following manner
\begin{align}
    A_{\mathrm{ref}}(\eta_{\mathrm{ta}})&=\dfrac{A}{24\pi\sqrt{3}}(1-f)^{-1/2}x_{\mathrm{ta}}^2\dfrac{\sigma_1^{11}G(\eta_{\mathrm{ta}})}{\sigma_0^{6}\sigma_2^{5}k_{\mathrm{max}}}\nonumber\\
    &=\Tilde{A}_{\mathrm{ref}}\sigma_0(1-f)^{-1/2}.
\end{align}
It is evident from the above expression that the interesting features of the reference spin at the turn around point reside in $\Tilde{A}_{\mathrm{ref}}$. In Fig.~\ref{Areftilde}, we have shown the behavior of $\Tilde{A}_{\mathrm{ref}}$.
It can be seen from the left panel of the figure that for $j\lesssim 2$, $\tilde{A}_{\mathrm{ref}}$ depends on $r_k$, where as for $j > 2$, it has no dependence on $r_k$. The same conclusion can be drawn from the right panel as well. This behavior is due to the fact that $\tilde{A}_{\mathrm{ref}}$ has terms which are of the form $r_k^{-j}$ both in its numerator and denominator. Therefore, for larger $j$ values, it loses its dependence on $r_k$.

\begin{figure}[H]
\centering
\includegraphics[scale=0.4]{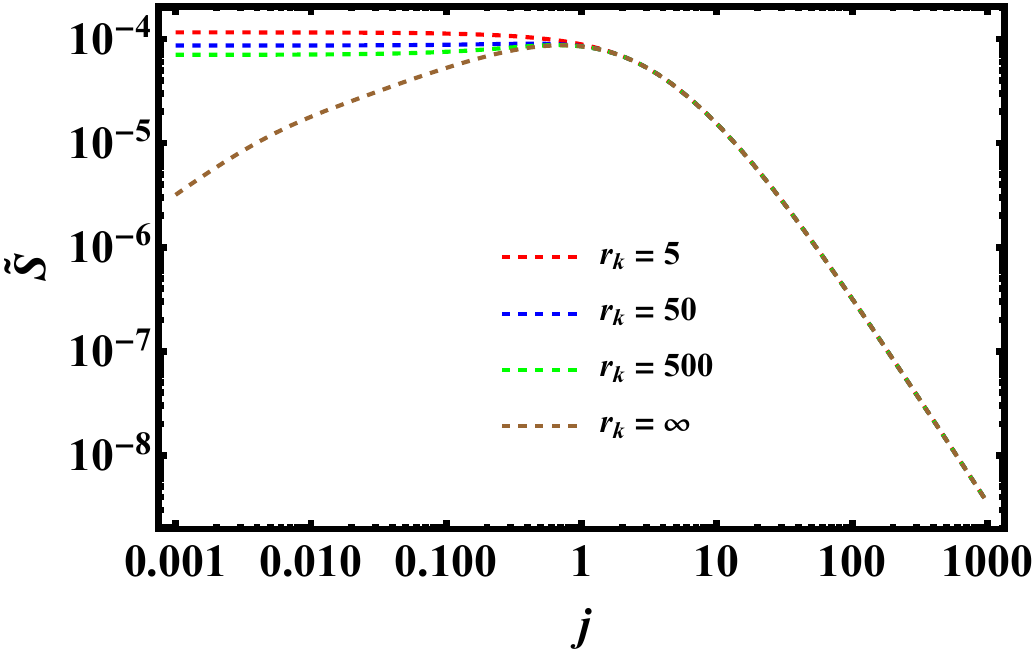}\\
\includegraphics[scale=0.38]{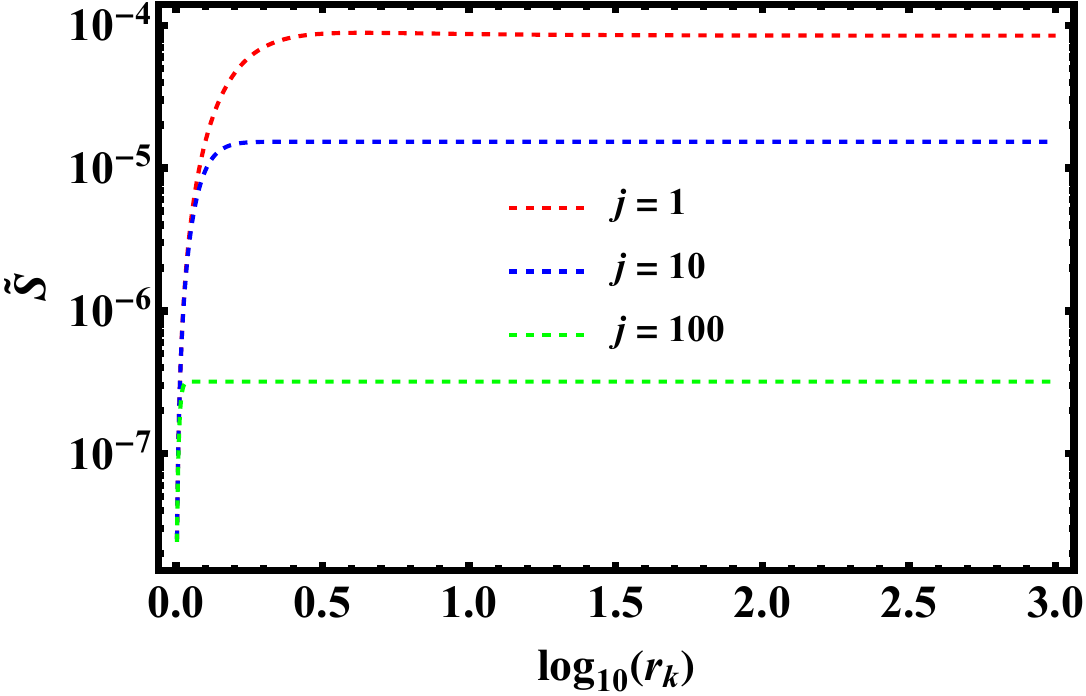}~~~~
\includegraphics[scale=0.4]{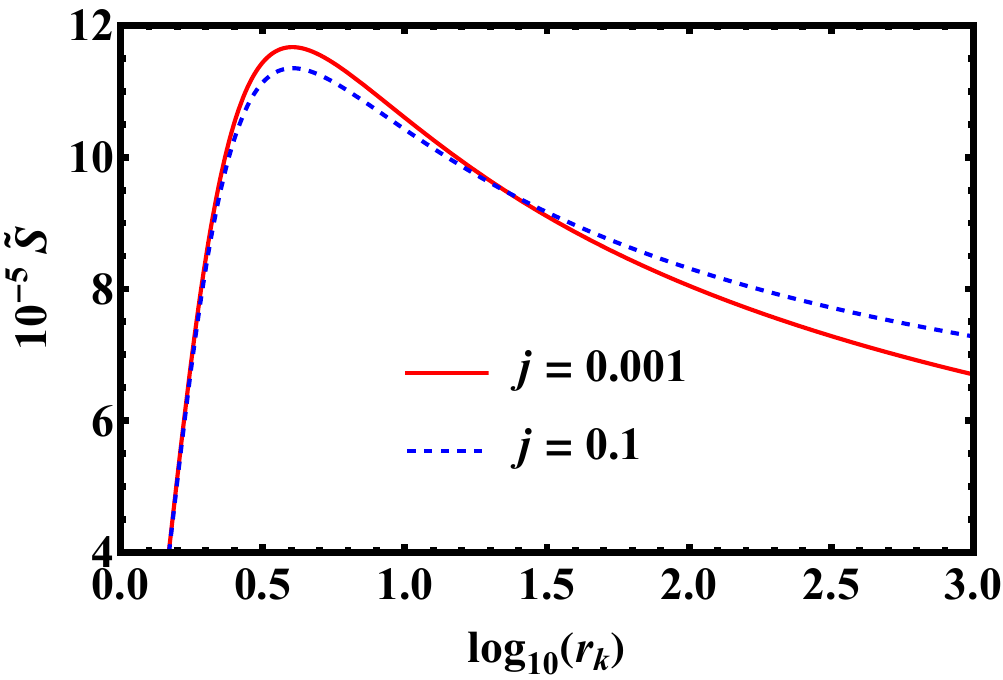}
\caption{(Top) The dependence of $\tilde{S}$ on $j$ with a few different $r_{k}$ values. (Bottom) The dependence of $\tilde{S}$ on $r_k$ for values of large $j$ (left) and $j\leq 0.1$ (right).}
\label{Stilde}
\end{figure}

As mentioned in Sec~\ref{sec:basdef}, we explain the RMS spin as
\begin{align}
\sqrt{\langle a_*^2 \rangle} = A_{\mathrm{ref}}\times 5.96\times \dfrac{\sqrt{1-\gamma^2}}{\gamma^6\nu},
\end{align}
where 
\begin{align}
\nu &= \dfrac{5}{2}\dfrac{\bar{\delta}_H}{\sigma_0},\nonumber\\
\gamma &= \dfrac{\sigma_1^2}{\sigma_0\sigma_2}. 
\end{align}
Here, $\gamma$ is a measure of monochromaticity, i.e., for completely monochromatic perturbation ($j\rightarrow\infty$ or $r_k\rightarrow 1$), $\gamma\rightarrow 1$. Therefore, we can express the RMS spin as
\begin{align}
\sqrt{\langle a_*^2\rangle} = \Tilde{S}\left(\dfrac{M_{\mathrm{PBH}}}{M_H}\right)^{-1/3}\left(\dfrac{\nu}{8}\right)^{-2},
\end{align}
where
\begin{align}
\Tilde{S}&=0.187\times \Tilde{A}_{\mathrm{ref}}\dfrac{\sqrt{1-\gamma^2}}{\gamma^6},\nonumber\\
M_{\mathrm{PBH}}&=1.14(1-f)^{3/2}M_{H}.
\end{align}
Here, $M_{\mathrm{PBH}}$ and $M_{H}$ are the PBH and the horizon mass, respectively. In Fig.~\ref{Stilde}, we have shown the dependence of $\Tilde{S}$ on $j$ and $r_k$.
It can be seen from the top panel that for $j\lesssim 0.1$, $\tilde{S}$ depends significantly on $r_k$, however for $j>0.5$, it loses its dependence on $r_k$. To illustrate this further, we have shown in the bottom left panel the dependence of $\tilde{S}$ on $r_k$ for a few cases where $j>0.5$. It can be seen that for $r_k> 2.5$, $\tilde{S}$ has no dependence on $r_k$. However, as shown in the bottom right panel, for smaller $j$ values, $\tilde{S}$ depends on $r_k$. This is due to the fact that similar to $\tilde{A}_{\mathrm{ref}}$, $\tilde{S}$ also consists of terms of the form of $r_k^{-j}$. However, due to the presence of $\gamma$, $\tilde{S}$ has even more power law dependence on $j$. As a result, unless the $j$ value is very small (nearly scale-invariant perturbation), the spin does not have any dependence on $r_k$. Therefore, in the following section, we divide the parameter space into two parts, i.e., (i) where we consider large $j$ values ($j> 0.1$) and keep $r_k$ as infinite and (ii) where we consider nearly scale invariant scenario ($j\lesssim 0.1$) and take finite values of $r_k$.
\section{Results}
\label{sec:res}
\begin{figure}[H]
\centering
\includegraphics[scale=0.63]{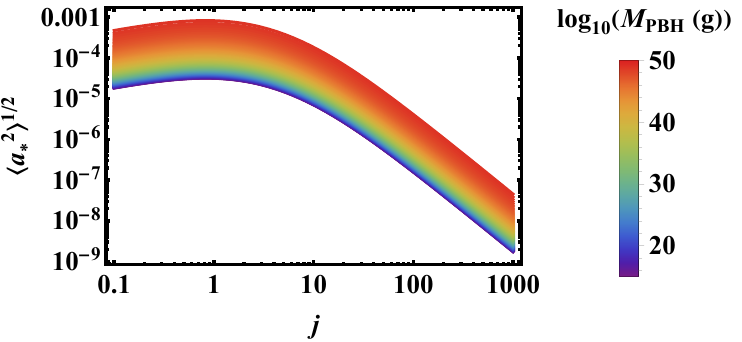}
\includegraphics[scale=0.63]{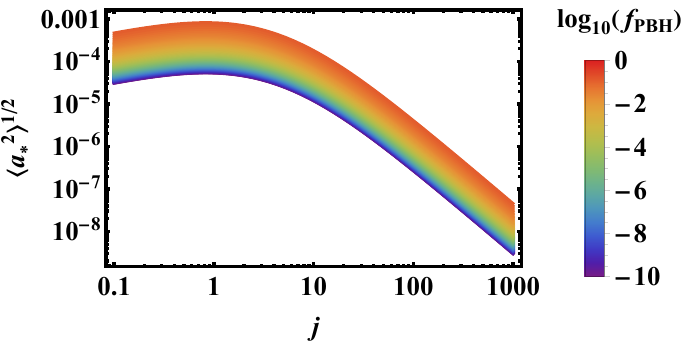}
\caption{The dependence of $\sqrt{\langle a_*^2\rangle}$ on $j$ for (Left) different values of $M_{\mathrm{PBH}}$ and $f_{\mathrm{PBH}}=1$ and (Right) different values of $f_{\mathrm{PBH}}$ and $M_{\mathrm{PBH}}=10^{50}\mathrm{g}$. In both the cases $\Omega_{\mathrm{CDM}}=0.26$.}
\label{astar1}
\end{figure}

In the previous section, we have shown the mathematical properties of the RMS spin of a PBH population in the index $j$ and width parameter $r_k$. In this section, we show the dependence of the PBH spin on $j$, $r_k$, PBH mass and abundance. In order to do that, we first consider the Press-Schechter formalism, which gives the fraction of the universe that collapses at the time of PBH formation as~\cite{Carr:1975qj}
\begin{align}
\beta_0=\sqrt{\dfrac{2}{\pi}}\dfrac{1}{\nu_{\mathrm{th}}}e^{-\nu_{\mathrm{th}}^2/2}=\sqrt{\dfrac{2}{\pi}}\dfrac{2}{5}\dfrac{\sigma_H}{\bar{\delta}_{H,\mathrm{th}}}\exp\left[-\left(\dfrac{5}{2}\right)^2\dfrac{\bar{\delta^2}_{H,\mathrm{th}}}{2\sigma_{H}}\right].
\end{align}
It is to be noted here that in incorporating the abundance and the mass into the expression of the RMS spin, we identify $\nu_{\mathrm{th}}$ as $\nu$, and $\sigma_H$ and $\sigma_0$. Furthermore, the abundance of the PBH population can be expressed as~\cite{Carr:2009jm}
\begin{align}
f_{\mathrm{PBH}}\approx 10^{18}\dfrac{\beta_0}{\Omega_{\mathrm{CDM}}}\left(\dfrac{M}{10^{15}\mathrm{~g}}\right)^{-1/2},
\end{align}
where $\Omega_{\mathrm{CDM}}$ is the current density of the dark matter of the universe. Now we can express the RMS spin as a function of both PBH mass and abundance as
\begin{align}
\sqrt{\langle a_*^2\rangle} = \dfrac{17.361\Tilde{S}\left(\dfrac{M}{M_H}\right)^{-1/3}}{23.484-1.25\log_{10}(f_{\mathrm{PBH}})-1.25\log_{10}\left(\dfrac{\Omega_{\mathrm{CDM}}}{0.26}\right)-0.625\log_{10}\left(\dfrac{M}{10^{15}\mathrm{~g}}\right)}.
\label{astarfinal}
\end{align}
Using Eq.~\eqref{astarfinal}, we can understand the behavior of the PBH spin. In Fig.~\ref{astar1}, we show the dependence of the PBH spin for $j>0.1$ on $j$, $M_{\mathrm{PBH}}$ and $f_{\mathrm{PBH}}$, while considering $r_k\rightarrow\infty$. 
It can be seen from Fig~\ref{astar1} that the behavior of $\sqrt{\langle a_*^2\rangle}$ on $j$ is similar to that of $\Tilde{S}$. However, it can also be seen that for fixed $f_{\mathrm{PBH}}$ ($M_{\mathrm{PBH}}$), $\sqrt{\langle a_*^2\rangle}$ increases with $M_{\mathrm{PBH}}$ ($f_{\mathrm{PBH}}$). In the parameter space we consider, the highest value of $\sqrt{\langle a_*^2\rangle}$ is $\sim 10^{-3}$ for $j\sim 1$, $M_{\mathrm{PBH}}=10^{50}\mathrm{~g}$ and $f_{\mathrm{PBH}}=1$.

\begin{figure}[H]
\centering
\includegraphics[scale=0.63]{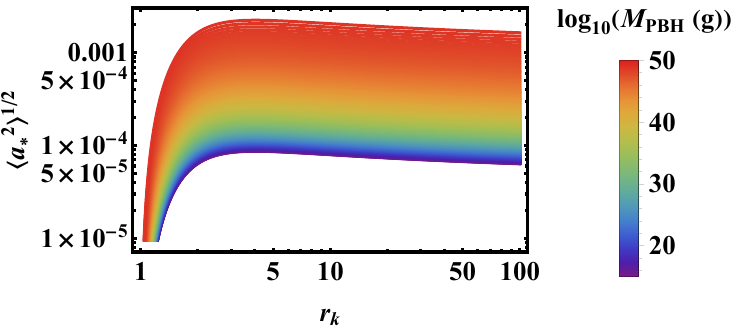}
\includegraphics[scale=0.63]{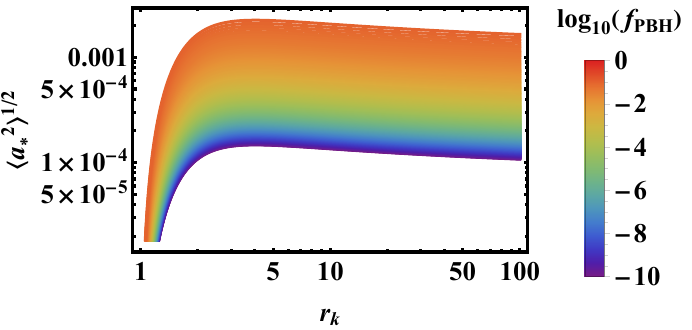}
\caption{The dependence of $\sqrt{\langle a_*^2\rangle}$ on $r_k$ for (Left) different values of $M_{\mathrm{PBH}}$ and $f_{\mathrm{PBH}}=1$ and (Right) different values of $f_{\mathrm{PBH}}$ and $M_{\mathrm{PBH}}=10^{50}\mathrm{g}$. For both the cases $j=0.1$ and $\Omega_{\mathrm{CDM}}=0.26$ are chosen.}
\label{astar2}
\end{figure}

On the other hand for $j\leq 0.1$, we have shown the behavior of $\sqrt{\langle a_*^2\rangle}$ on $r_k$, $M_{\mathrm{PBH}}$ and $f_{\mathrm{PBH}}$ in Fig.~\ref{astar2}. It is worth mentioning here that since in the previous section we found that for $j\leq 0.1$, $\Tilde{S}$ does not have a strong dependence on $j$, we have kept $j=0.1$ for the figure below. 
It is evident from the above figure that as in the previous case, here the behavior of $\sqrt{\langle a_*^2\rangle}$ on $r_k$ is similar to that of $\Tilde{S}$. Furthermore, in this case as well, $\sqrt{\langle a_*^2\rangle}$ increases with both $M_{\mathrm{PBH}}$ and $f_{\mathrm{PBH}}$. In this case for the parameter space, we find the highest value of $\sqrt{\langle a_*^2\rangle}$ is $\sim 2.5\times 10^{-3}$ for $r_k\sim 3.5$, $M_{\mathrm{PBH}}=10^{50}\mathrm{~g}$ and $f_{\mathrm{PBH}}=1$. It is also to be noted here that even for $r_k>3.5$, the rate at which $\sqrt{\langle a_*^2\rangle}$ decreases is very low, i.e., even for $r_k = 100$, we obtain $\sqrt{\langle a_*^2\rangle}\sim 1.6\times 10^{-3}$ for $M_{\mathrm{PBH}}=10^{50}\mathrm{~g}$ and $f_{\mathrm{PBH}}=1$, which is larger than what we obtain for $j>0.1$. Therefore, we conclude that for a broad power spectrum the highest value of the initial spin of PBHs occurs when the PBHs are created from a nearly scale invariant power spectrum with a value of $r_k\sim 3.5$.

It would be useful to summarize the results for the mass ranges of more physical interest. Irrespective of the mass range, 
the largest RMS of $a_{*}$ is realized for a nearly scale invariant power spectrum $(j\lesssim 0.1)$ with $r_{k}\sim 3.5$.   
The upper limit value is 
$\sim 1\times 10^{-4}$ for $M=10^{17}-10^{23}$ g with $f_{\rm PBH}=1$, 
 $\sim 1.7\times 10^{-4}$ for $M=1-100 M_{\odot}$ with $f_{\rm PBH}=0.1$ and 
$\sim 2.5\times 10^{-3}$ for an incredibly massive case 
of $M=10^{50}$ g with $f_{\rm PBH}=1$. The RMS only weakly depends on $f_{\rm PBH}$ and the smaller the $f_{\rm PBH}$ is, the smaller the RMS becomes.
\section{Summary and Conclusion}
\label{sec:concl}
In this article, we perform a study to obtain the initial spin of PBHs created during the radiation dominated phase of the universe from a broad curvature power spectrum.

We have considered a power law shape for the curvature perturbation, where the we have used the power law index ($j$) and the width parameter ($r_k$) to indicate how broad the power spectrum is, i.e., if the power law index is zero and the width parameter in infinity, that indicates a scale-invariant power spectrum, where as very high values of the power law index (or width parameter values close to unity) point towards a monochromatic power spectrum. In order to approach toward the calculation of the spin, we first obtained the expressions for spectral index and the shape of the profile of the perturbation. In this regard, it is to be mentioned here that in the calculations, we had only considered the contributions from the length scale equivalent to the horizon length or larger. Furthermore, for the power law index values approaching zero, we have considered an upper value for the length scale that contributes, i.e., we have bounded the width parameter to finite values. Next we obtained the turn around point, i.e., the point where the region, which eventually collapses into a black hole, decouples from the background. We find that irrespective of the shape of the perturbation, the value of the turn around point remains constant as long as we consider radiation domination. 

After calculating the turn around, we obtained the reference and the RMS spin and their dependence on the power law index and the width parameter. We found that for the cases where the power law index is large, the reference spin increases with and the RMS spin decreases with increasing values of the index. For the cases where the power law index is vanishingly small, we found that the reference spin decreases as we increase the width parameter, whereas the RMS spin initially increases and then decreases for the same. Finally we used the estimate from the Press-Schechter formalism to express the zeroth spectral moment ($\sigma_0$) in terms of the PBH mass and abundance in order to gain insight into the dependence of the RMS spin on those quantities. We found that for both large and small values of the power law index, the spin of PBH increases with mass and abundance of the PBH. In this regard, it is to be noted that although this work is in the same spirit of that of Refs.~\cite{DeLuca:2019buf, Harada:2020pzb}, in this work we have considered the a general form of a curvature power spectrum motivated from different early universe phenomenon. Furthermore, we have also considered the deviation from monochromaticity in the form of the quantity $\gamma$ and incorporated its effect as opposed to considering benchmark values. This showed that the narrowness of a curvature power spectrum eventually ends up suppressing the spin. We further understood that highest value of initial PBH spin is nearly $2.5\times 10^{-3}$ and it arises from PBHs generating from nearly scale invariant perturbation where the width parameter is close to 3.5.  

It is to be noted that in this work, we have considered a few approximations, such as we have divided our parameter space into two parts, i.e., $j>0.1$ and $j\leq 0.1$. Furthermore, we have considered the Press-Schechter formalism to obtain the influence of PBH mass and abundance on the initial spin of the PBH, whereas a more detailed analysis based on peak theory would have lead to a more refined dependence. However, for the purpose of our work, i.e., to understand the overall influence of PBH mass and abundance on spin, the Press-Schechter formalism is sufficient. We have also shown the PBH spin only for the case of PBH mass close to the horizon mass, whereas it is known that in case the PBH mass is much lower than the horizon mass, the spin is enhanced, which can be seen from our expression for the PBH spin as well. However, since PBH mass much less than the horizon mass suppresses the abundance, we did not consider it in our work. Finally, we found that the maximum spin for the cases we have considered in this study is $\sim 2.5\times 10^{-3}$, which is very small. However, some epoch of the universe with softer equations of state may give rise to much higher spin, which we leave for future work.

In conclusion, we believe that our work leads to better understanding of initial PBH spin for PBHs created from broad curvature power spectrum as there are many early universe events such as inflation, first order phase transitions, etc, which leads to broad power spectrum. However, further studies are important to obtain the spin in each of these specific cases to understand the PBH spins. 

%
 
\acknowledgments
IKB acknowledges the support by the MHRD, Government of India, under the Prime Minister's Research Fellows (PMRF) Scheme, 2022. IKB thanks Tanmoy Kumar, Nandini Das, and Shreecheta Chowdhury for useful discussions.
TH was supported by JSPS KAKENHI Grant Numbers 20H05853 and 24K07027. TH is grateful to CENTRA, Departamento de F{í}sica,
Instituto Superior T\'{e}cnico -- IST at Universidade de Lisboa, and
Niels Bohr International Academy at Niels Bohr Institute
for their hospitality during the writing of the manuscript.

\bibliographystyle{JHEP}
\bibliography{pbhspinbps_ref.bib}

\end{document}